\documentclass{article}
\usepackage{amsmath,amsthm}
\usepackage{graphicx,epstopdf,epsfig,multirow,epic,bm}
\usepackage{lineno,hyperref}
\usepackage{amsfonts}
\usepackage{subfigure}
\usepackage{amssymb}
\usepackage{algorithm}
\usepackage{algpseudocode}

\normalsize \rm
\begin{document}

\begin{center}
{\Large  \textbf { OPTT: Optimal Piecewise Transformation Technique for Analyzing Numerical Data under Local Differential Privacy }}\\[12pt]
{\large Fei Ma$^{a,}$\footnote{~The author's E-mail: mafei123987@163.com. },\quad Renbo Zhu$^{c,}$, \quad  Ping Wang$^{b,c,d,}$\footnote{~The author's E-mail: pwang@pku.edu.cn.} }\\[6pt]
{\footnotesize $^{a}$ School of Electronics Engineering and Computer Science, Peking University, Beijing 100871, China\\
$^{b}$ National Engineering Research Center for Software Engineering, Peking University, Beijing, China\\
$^{c}$ School of Software and Microelectronics, Peking University, Beijing  102600, China\\
$^{d}$ Key Laboratory of High Confidence Software Technologies (PKU), Ministry of Education, Beijing, China}\\[12pt]
\end{center}

\begin{quote}
\textbf{Abstract:} Privacy preserving data analysis (PPDA) has received increasing attention due to a great variety of applications. Local differential privacy (LDP), as an emerging standard that is suitable for PPDA, has been widely deployed into various real-world scenarios to analyze massive data while protecting against many forms of privacy breach. In this study, we are mainly concerned with piecewise transformation technique (PTT) for analyzing numerical data under local differential privacy. We provide a principled framework for PTT in the context of LDP, based on which PTT is studied systematically. As a result, we show that (1) many members in PTTs are asymptotically optimal when used to obtain an unbiased estimator for mean of numerical data, and (2) for a given privacy budget, there is PTT that reaches the theoretical low bound with respect to variance. Next, we prove by studying two classes of PTTs in detail that (1) there do not exist optimal PTTs compared to the well-used technique, i.e., Duchi's scheme, in terms of the consistency noisy variance, (2) on the other hand, one has the ability to find a great number of PTTs that are consistently more optimal than the latter with regard to the worst-case noisy variance, which is never reported so far. When we are restricted to consider only the high privacy level, enough PTTs turn out to be optimal than the well-known Laplace mechanism. Lastly, we prove that for a family of PTTs, the correspondingly theoretical low bound of noisy variance follows $O(\epsilon^{-2})$ when considering the high privacy level.
\\

\textbf{Keywords:} Piecewise Transformation Technique, Local differential private mechanism, Optimization \\

\end{quote}

\section{Introduction}

Differential privacy (DP for short), firstly developed by Dwork et al \cite{Dwork-2006,Dwork-2006-1}, has been accepted as a de faco standard for privacy preserving data analysis according to its own merits. Specifically speaking, DP provides a provable privacy protection in the jargon of mathematics regardless of the adversary's background knowledge. In the setting of DP, there always exists a trusted party who is assumed to not steal or leak any private information of individual data owner in the centralized case \cite{Dwork-2006-1}. More generally, the party serves as a data curator (or, collector) who executes a randomized mechanism on raw data collected from each user and reports the sanitized results for a given query. However, it is well known that such an assumption is not always reasonable in real-world setting \cite{Imola-2021}-\cite{Pastore-2021}. The reason for this is that with the rapid development of data mining and analysis, data is now deemed as the crucial assets in various kinds of fields. Data owners themselves become more and more concerned with their own data, and worry that they no longer take control of data as long as raw data are sent to that so-called trusted party. As a result, they are reluctant to submit raw data to the curator. To address this concern, the local version related to differential privacy has been put forward \cite{Duchi-2013}, which is viewed as Local differential privacy (LDP for short).

In the last several years, LDP has received more attention from academic. In theory, Duchi et al have analyzed the fundamental properties of LDP in more detail \cite{Duchi-2014}-\cite{Duchi-2018}. Accordingly, they came up with some simple protocols that have been proven useful in various applications \cite{Wang-2019-ICDE}-\cite{Gu-2020}. For instance, in view of the simplicity, the scheme proposed in \cite{Duchi-2014} (which is discussed in detail later) has been widely used as the building-block to design more complicated and practical protocols for numerical data \cite{Wang-2019-ICDE} and Key-vale data \cite{Gu-2020}. In addition, many other previous schemes, which are established for the purpose of satisfying DP, can also be straightforwardly extended to meet the criteria in the scenario of LDP. Consequently, they are selected to serve as ingredients for new LDP-protocols \cite{Wang-2019-ICDE}-\cite{Gu-2020}. The most famous among them are the classic Laplace mechanism \cite{Dwork-2006-1} and Randomized Response \cite{Warner-1965}. As well known, these two LDP-schemes have been deployed into a great variety of applications \cite{Li-2019}-\cite{Josep-2020} due to their own celebrated structural simplicity.  Therefore, it is of great interest to build up novel and simple LDP-schemes that can be able to act as building-blocks for more complex protocols in the field of LDP. What's more, it is more interesting and challenging to design optimal schemes compared to previous ones with respect to some widely-used metrics including variance.

It is well known that in prior work, LDP-based protocols have been used to analyze various types of analysis tasks, for instance, mean estimation of numerical data \cite{Wang-2019-ICDE}, frequency/histogram estimation on categorical data \cite{Wang-2017-USNIX} and discovering heavy hitter \cite{Qin-2016}. In recent years, there are more attention paid on other complicated queries including computing mean value over a single numeric attribute of multidimensional data \cite{Duchi-2018}, frequent item/itemset mining of itemset data \cite{Wang-2018-SP}, generating synthetic social graphs \cite{Wei-2020} and counting subgraphs on graph \cite{Imola-2021}. Nonetheless, some simple LDP-schemes, which are used as ingredients in protocols mentioned above, are not be optimal in terms of many metrics such as variance. In other words, it is of great interest to either design novel schemes in order to optimize some previous protocols or analyze in detail those published protocols for the goal of finding theoretical bounds.

Last but not the least, there are in fact many LDP-protocols deployed into a wide range of real-world scenarios, for example, Apple's iOS, macOS and Safari \cite{Apple-2016}, Microsoft Windows 10 \cite{Ding-2017}, Google Chrome software systems \cite{Erlingsson-2014}, and Alibaba \cite{Wang-2019-MOD}. Specifically, Google has employed RAPPOR in Chrome so as to collect web browsing behavior while providing LDP guarantees \cite{Erlingsson-2014}. Apple has already also leveraged LDP-based mechanisms to uncover popular emojis, popular health data types, and media playback preference in Safari \cite{Apple-2016}. This triggers the study of LDP protocols. In fact, these practical protocols are based on some simple schemes that are initially designed for specific tasks, such as mean estimation and distribution in numerical data \cite{Wang-2019-ICDE,Wang-2019-TPDS}, heavy hitter discovery and frequency estimation in categorial data \cite{Wang-2017-USNIX,Qin-2016}, and so on. In a nutshell, it is important to build up simple schemes for something very specific in the context of LDP.

Motivated by this, we propose a novel and simple scheme for a fundamental query, i.e., mean value estimation over a single numeric attribute of multidimensional data, in this paper. In the meantime, the proposed scheme is analyzed in depth and, at the same time, is compared with two typical schemes. To summarize, the main contribution of this paper is shown as below.

\emph{Contribution}---We propose a principled framework for PTT in the context of LDP, and then study PTTs systematically. We show that (i) there are many members in PTTs that are asymptotically optimal when adopted to obtain an unbiased estimator for mean of numerical data, and (ii) for a given privacy budget, one designs optimal PTT that reaches the theoretical low bound with respect to noisy variance. We declare that when analyzing numerical data under LDP, there are no optimal PTTs compared to Duchi's scheme in terms of the consistency noisy variance. On the other hand, we have the ability to find a great number of PTTs that are consistently more optimal than the latter with regard to the worst-case noisy variance. We also certify that there are always optimal PTTs contrasted with the classic Laplace mechanism at the high privacy level. More importantly, we prove that for a family of PTTs, the correspondingly theoretical low bound of noisy variance obeys $O(\epsilon^{-2})$ as privacy budget $\epsilon\rightarrow 0$.

\emph{Roadmap}---The rest of this paper is organized as follows. Section II introduces some basic notations used later. Section III first revisits two well-known LDP protocols, namely, the classic Laplace mechanism and Duchi's scheme, and then presents our main results. Section IV briefly reviews the related work. Finally, we close this paper in Section V.

\section{Notations and Definitions}

Here, we will introduce some fundamental concepts and notations used throughout this paper, including LDP and graph, First of all, let us bring the concrete definition of LDP \cite{Dwork-2006}.

\textbf{Definition 1 \cite{Duchi-2013}} A randomized mechanism $\mathcal{M}$ turns out to satisfy \emph{$\epsilon$-local differential privacy}, shorted as $\epsilon$-LDP, if and only if for arbitrary two inputs $D_{i}$ and $D_{j}$ in domain $\mathcal{D}$ and for an arbitrary output $A^{\ast}$ in range $\mathcal{A}$, the following inequality always holds

\begin{equation}\label{eqa:MF-2-1}
\mathrm{Pr}[\mathcal{M}(D_{i})=A^{\ast}]\leq e^{\epsilon}\mathrm{Pr}[\mathcal{M}(D_{j})=A^{\ast}]
\end{equation}
where $\mathrm{Pr}[\cdot]$ represents probability and parameter $\epsilon$ is often referred to as privacy budget that measures the level of privacy-preserving of mechanism $\mathcal{M}$.

By definition, $\epsilon$-LDP mechanism means that after observing the output $A^{\ast}$, an arbitrary adversary including data collector cannot infer whether the input is $D_{i}$ or $D_{j}$ with high confidence controlled by parameter $\epsilon$. That is to say, this kind of mechanisms provide data owner with plausible deniability. A smaller $\epsilon$ corresponds to a stronger privacy-preserving guarantee.

Perhaps, one of the most important merits of DP is the so-called \emph{Sequential Composition}. In the following, we introduce concrete demonstration associated with $\epsilon$-LDP.

\textbf{Definition 2 \cite{Dwork-2014}} Given a collection of randomized mechanisms $\mathcal{M}_{i}$ ($i\in\{1,2,\dots,n\}$), each of which satisfies $\epsilon_{i}$-local differential privacy. Then, the sequence of mechanisms $\mathcal{M}_{i}$ together provides $\Sigma_{i=1}^{n}\epsilon_{i}$-local differential privacy.

It is worth noting that the abovementioned property clearly suggests that one has the ability to build up more complicated differential privacy mechanisms based on some simple and basic building-blocks. For instance, given a privacy budget $\epsilon$, we can partition it into several smaller portions such that each portion is used for one randomized mechanism to give a privacy guarantee. On the other hand, the elegant property often ensures that the resulting mechanism is suboptimal. Therefore, more effort has been done in order to make improvement on mechanisms of such type in the last years \cite{Kairouz-2017}. See \cite{Kairouz-2017} for more information.

In the rest of this section, let us state the main problem to be considered. Precisely, we aim to discuss one type of analysis task under $\epsilon$-LDP, i.e., mean value and frequency estimation on numerical data. In fact, this is a hot topic that has been widely studied, and thus a number of schemes have been proposed in the literature of DP/LDP \cite{Duchi-2014,Wang-2019-ICDE,Nguyen-2016,Wang-2017-USNIX}. Formally, we showcase the corresponding definition as below.

\textbf{Definition 3} Given a set of data owners $o_{i}$ ($i\in \{1,2,\dots,n\}$), each owner $o_{i}$ has an attribute $A_{i}$. Without loss of generality, we suppose that the domain of attributes $A_{i}$ is $[-1,1]$\footnote[1]{An arbitrary interval $[a,b]$ is easy to convert into the standard form $[-1,1]$ using the following manipulation: For each element $u$ in $[a,b]$, one needs to calculate a quantity $u'=\frac{2}{b-a}u-\frac{b+a}{b-a}$. Consequently, all fresh elements $u'$ belong to $[-1,1]$ as desired. Bear in mind that such a simple transformation technique enables us to gain many surprising results as will be shown in the following sections.}. Then, mean $\langle A\rangle$ associated with a given attribute $A$ is defined as

\begin{equation}\label{eqa:MF-2-3-1}
\langle A\rangle=\frac{1}{n}\sum_{i=1}^{n}A_{i}\delta_{A_{i},A}
\end{equation}
where symbol $\delta_{i,j}$ is the Kronecker delta function in which $\delta_{i,j}$ is equal to $1$ when $i=j$, and $0$ otherwise. More generally, each owner $o_{i}$ possesses number $d_{i}$ of distinct attributes $A_{i,l}$. And then, for a given attribute $A$, mean $\langle A\rangle$ is given by

\begin{equation}\label{eqa:MF-2-3-2}
\langle A\rangle=\frac{1}{n}\sum_{i=1}^{n}\sum_{l=1}^{d_{i}}A_{i,l}\delta_{A_{i,l},A}.
\end{equation}

\section{Main schemes for numerical attribute}

The goal of this section is to analyze in detail single numerical attribute in a manner following $\epsilon$-LDP constraint. More specifically, we will provide a scheme, which turns out to satisfy $\epsilon$-LDP, for PPDA in the setting where each data owner $o_{i}$ has a unique attribute $A_{i}$ in $[-1,1]$. At the high level, the heart of the proposed scheme is to inject noise drawn from some distribution into attribute $A_{i}$ for the purpose of creating noisy attribute $\widetilde{A_{i}}$, which is similar in spirit to the well-known Laplace mechanism \cite{Dwork-2006-1}. In fact, there have been many similar mechanisms developed based on such type of thought, such as Gaussian mechanisms \cite{Yang-2021}. Besides that, Duchi \emph{et al} put forward another novel mechanism, which is called Duchi's scheme hereafter for our purpose, in order to address the issue mentioned above when required to follow $\epsilon$-LDP \cite{Duchi-2018}. It should be noted that Duchi's scheme, as one basic building-block, has been widely used to create a great number of protocols for various applications \cite{Nguyen-2016}-\cite{Gu-2020}. Note also that before beginning with our discussions, two famous mechanisms, i.e., Laplace mechanism and Duchi's scheme, need to be reviewed. The reason for this is twofold: (1) As known, these two mechanisms have been referred to as the fundamental building-blocks in the literature of PPDA due to their own simplicity; (2) In this paper, our goal is to propose a new and simple mechanism that is shown to produce optimal result utility compared with these two well-known mechanisms when provided some constraints. Additionally, we would like to see that our mechanism can have the potential to serve as one fundamental building-block suitable for $\epsilon$-LDP protocols in the future.

\subsection{Laplace mechanism and fundamental properties}

In the pioneering work due to Dwork \cite{Dwork-2006-1}, \emph{Laplace mechanism }is used to enforce differential privacy in the LDP scenario. The concrete implementation of Laplace mechanism is as follows: Given an attribute $A_{i}$, the corresponding data owner $o_{i}$ samples a sample $\mathfrak{L}(2/\epsilon)$ from Laplace distribution with scale $\lambda=2/\epsilon$ at random, and then insert $\mathfrak{L}(2/\epsilon)$ as noise to $A_{i}$ to obtain $\widetilde{A_{i}}=A_{i}+\mathfrak{L}(2/\epsilon)$. After that, the noisy attribute $\widetilde{A_{i}}$ is submitted to data curator for the purpose of making some simple statistic analysis, such as mean estimation.  The previous work has shown that Laplace mechanism is subject to the following statements in lemma 1.

\textbf{Lemma 1 \cite{Dwork-2014}}  When implementing Laplace mechanism for an arbitrary entry $A_{i}$ to yield an estimation $\widetilde{A_{i}}$, the corresponding mean $\mathbb{E}_{\mathfrak{L}}[\widetilde{A_{i}}]$ and variance $\mathrm{Var}_{\mathfrak{L}}[\widetilde{A_{i}}]$ follow

\begin{equation}\label{eqa:MF-3-1}
\mathbb{E}_{\mathfrak{L}}[\widetilde{A_{i}}]=A_{i},\quad \mathrm{Var}_{\mathfrak{L}}(\widetilde{A_{i}})=\frac{8}{\epsilon^{2}}.
\end{equation}

From Eq.(\ref{eqa:MF-3-1}), one can see that the obtained estimation $\widetilde{A_{i}}$ is unbiased. At the same time, it is easy to see that variance $\mathrm{Var}_{\mathfrak{L}}[\widetilde{A_{i}}]$ is independent of input $A_{i}$. Also, it only depends on Laplace distribution $\mathfrak{L}(2/\epsilon)$.  After each data owner $o_{i}$ runs Laplace mechanism, the data curator collects all the estimations $\widetilde{A_{i}}$, and then makes use of $\sum_{i=1}^{n}\widetilde{A_{i}}/n$ to exactly represent mean over dataset what he/she is interested in. Note that doing so leads to an error scale $O(\epsilon^{-1}n^{-1/2})$. As will see later, this is a relatively higher error than that behaved by our scheme (which is detailed later) when discussing high privacy level, i.e., $\epsilon\in(0,1]$.

Since then, a large number of mechanisms satisfying $\epsilon$-DP/$\epsilon$-LDP have been established in a similar manner used in the development of Laplace mechanism \cite{Dwork-2014}. From the methodology point of view, mechanisms of such kind are established based on the following principled framework:

\emph{\textbf{Framework 1}: Given an attribute $A_{i}$, data owner $o_{i}$ needs to seek for a distribution, based on which the noise required to inject into $A_{i}$ is drawn. And then, all the obtained results should be shown to comply with requirement in Def.1.}

Clearly, such mechanisms can be viewed as a transform from raw data into continue variables. It is worth mentioning that in some cases, the sanitized results from these mechanisms may be biased when used to analyze some parameters such as mean. To addressed this issue, some pre-processing or post-processing techniques must be adopted. For example, given a random variable $x$, the standard Laplace distribution follows $\mathfrak{L}(\mu,b)=\frac{1}{2b}\exp(-\frac{|x-\mu|}{b})$ where $\mu$ and $b$ represent location parameter and scale parameter, respectively. In order to leverage Laplace distribution to build up the classic Laplace mechanism, one first modifies the standard version using pre-processing method, namely, setting $\mu=0$ and $b=2/\epsilon$ \cite{Dwork-2014}. Again for instance, both Soria-Comas \emph{et al} \cite{Soria-Comas-2013} and Geng \emph{et al} \cite{Geng-2015} take advantage of piece-wise constant probability distribution to construct $\epsilon$-LDP schemes for single numerical attribute. Very recently, a follow-up scheme called piecewise mechanism is proposed by Wang \emph{et al} \cite{Wang-2019-ICDE} using piecewise probability distribution, which falls into the principled framework mentioned above. In addition, Wang \emph{et al} have shown that the proposed piecewise mechanism is more optimal than the classic Laplace mechanism in some settings. We encourage interested readers to refer Ref.\cite{Wang-2019-ICDE} for more detailed information.

\subsection{Duchi's mechanism and fundamental properties}

In the following, we introduce another scheme suitable for analyzing numerical attribute that is presented by Duchi \emph{et al} \cite{Duchi-2018,Duchi-2013}. For convenience, we denote by \emph{Duchi's mechanism} this scheme. As opposed to the classic Laplace mechanism, the core of Duchi's mechanism is to convert a raw data into discrete variables. More precisely, a raw data is perturbed into two predefined values using Duchi's mechanism. The concrete procedure is shown in Algorithm 1.

\begin{algorithm}
	\caption{Duchi's mechanism for one-dimensional numeric attribute \cite{Duchi-2018}}
	\label{alg:Framwork}
	\begin{algorithmic}[1]
		\Require
		\quad attribute $A_{i}\in[-1,1]$;
		privacy budget $\epsilon$
\Ensure  attribute $\widetilde{A_{i}}\in\{\frac{e^{\epsilon}+1}{e^{\epsilon}-1},-\frac{e^{\epsilon}+1}{e^{\epsilon}-1}\}$
		\State Sample a Bernoulli variable $\omega$ such that
$$\mathrm{Pr}[\omega=1]=\frac{e^{\epsilon}-1}{2(e^{\epsilon}+1)}\times A_{i}+\frac{1}{2}$$
        \If {$\omega=1$}
					\State $\widetilde{A_{i}}=\frac{e^{\epsilon}+1}{e^{\epsilon}-1}$
\Else
\State $\widetilde{A_{i}}=-\frac{e^{\epsilon}+1}{e^{\epsilon}-1}$
				\EndIf
\State
		\Return $\widetilde{A_{i}}$
	\end{algorithmic}
\end{algorithm}

\textbf{Lemma 2 \cite{Duchi-2018}} When implementing Duchi's mechanism for an arbitrary attribute $A_{i}$ to yield an estimation $\widetilde{A_{i}}$, the corresponding mean $\mathbb{E}_{\mathfrak{D}}[\widetilde{A_{i}}]$ and variance $\mathrm{Var}_{\mathfrak{D}}(\widetilde{A_{i}})$ follow

\begin{equation}\label{eqa:MF-3-2}
\mathbb{E}_{\mathfrak{D}}[\widetilde{A_{i}}]=A_{i},\quad \mathrm{Var}_{\mathfrak{D}}(\widetilde{A_{i}})=\left(\frac{e^{\epsilon}+1}{e^{\epsilon}-1}\right)^{2}-A^{2}_{i},
\end{equation}
and Duchi's mechanism satisfies $\epsilon$-LDP.

As above, Duchi's mechanism produces unbiased estimation for attribute $A_{i}$ as well. Yet, the corresponding variance $\mathrm{Var}_{\mathfrak{D}}(\widetilde{A_{i}})$ is dependent on input $A_{i}$. This is completely different from Laplace mechanism. In addition, the data curator may also utilize $\sum_{i=1}^{n}\widetilde{A_{i}}/n$ to exactly denote mean over dataset what he/she cares about. Particularly, doing so causes an error scale $O(\epsilon^{-1}n^{-1/2})$ when privacy budget $\epsilon$ goes to zero, i.e., $\epsilon\rightarrow 0$, which is the same as the classic Laplace mechanism. In other words, two fundamental and important mechanisms introduced above behave similar performance as $\epsilon\rightarrow 0$ when utilized to yield mean estimation on dataset.

Along the same research line as Duchi's mechanism, many variants have been proposed and studied in the realm of PPDA. For example, Nguyen \emph{et al} establish the so-called \emph{Harmony} for analyzing numerical data in an $\epsilon$-LDP manner \cite{Nguyen-2016}. In a follow-up work \cite{Ye-2019}, Ye \emph{et al} indeed utilized \emph{Harmony} as an important ingredient to construct the \emph{PrivKVM} for analyzing Key-Value pair in an $\epsilon$-LDP way. More recently \cite{Gu-2020}, Gu \emph{et al} analyze the deficiencies planted on \emph{PrivKVM} and further develop a more optimal scheme, namely \emph{PCKV-UE}. It is worth noting that all the specific schemes are built on the basis of the thought behind Duchi's mechanism. To put this another way, Duchi's mechanism is a key ingredient for the previously published mechanisms \cite{Nguyen-2016}-\cite{Gu-2020}. It is natural to ask whether there are other available mechanisms showing optimal performance or not. Fortunately, there is positive answer to this issue as will be shown shortly.

Let us conclude this subsection by succinctly distinguishing Laplace mechanism and Duchi's mechanism. From Eqs.(\ref{eqa:MF-3-1}) and (\ref{eqa:MF-3-2}), it is clear to the eye that both mechanisms result in an unbiased estimation for an arbitrary raw data $A_{i}$. To make further progress, the mean estimations yielded from these two  mechanisms are also unbiased. On the other hand, the corresponding error scales behave distinctly. In theory, for a given attribute $A_{i}$, we can derive a unique solution $\epsilon_{i}$ to the coming equation

\begin{equation}\label{eqa:MF-3-2-1}
\left(\frac{e^{\epsilon}+1}{e^{\epsilon}-1}\right)^{2}-A^{2}_{i}-\frac{8}{\epsilon^{2}}=0,
\end{equation}
then say that for arbitrary $\epsilon\in(0,\epsilon_{i}]$, Duchi's mechanism is more optimal than Laplace mechanism with respect variance, and the opposite result can be obtained for arbitrary $\epsilon\in(\epsilon_{i},+\infty]$. To put this another way, it is benefit to adopt Duchi's mechanism to perturb raw data $A_{i}$ for a given privacy budget $\epsilon\leq\epsilon_{i}$. In essence, it is difficult to derive an exact solution of quantity $\epsilon_{i}$ in a mathematically rigorous mainly because Eq.(\ref{eqa:MF-3-2-1}) is a transcendental equation. To validate the correctness of our statements, we plot numerical solutions to Eq.(\ref{eqa:MF-3-2-1}) in Fig.1 where a series of distinct attributes $A_{i}$ are considered. See Fig.1 for more information.

\begin{figure}
\centering
  \includegraphics[height=6cm]{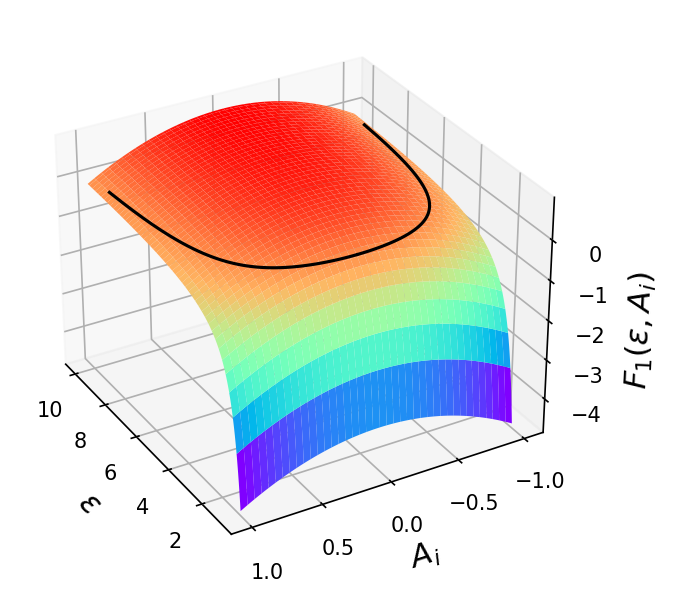}\\
{\small Fig.1. (Color online) The diagram of numerical solutions to Eq.(\ref{eqa:MF-3-2-1}) where we are limited in the setting of privacy budget $\epsilon\in(0,10]$. For brevity, we introduce a new function $F_{1}(\epsilon,A_{i})\triangleq\left(\frac{e^{\epsilon}+1}{e^{\epsilon}-1}\right)^{2}-A^{2}_{i}-\frac{8}{\epsilon^{2}}$. The black solid points represent solutions of Eq.(\ref{eqa:MF-3-2-1}). In this case, all black solid points are collected together to yield a black curve. Clearly, when focusing on the high privacy level, namely, $\epsilon\in(0,1]$, function $F_{1}(\epsilon,A_{i})$ is consistently smaller than zero, which implies that Duchi's mechanism is more optimal than the classic Laplace mechanism. Meanwhile, the optimality planted on Duchi's mechanism is considerably degraded with increasing $\epsilon$ from the high privacy level to a more broad spectrum.  }
\end{figure}

\subsection{Piecewise transformation technique}

In this subsection, we will provide a simple yet fundamental mechanism for analyzing numerical attribute in an $\epsilon$-LDP fashion which is referred to as piecewise transformation technique (PTT) for brevity. Before beginning with detailed descriptions about PTT, two main tips need to be highlighted. The first tip is that in principle, PTT is a kind of perturbation technique which is established in a similar manner as in Laplace mechanism. Specifically, it converts an attribute $A_{i}$ into a continue variable by adding noise drawn from a delicately designed probability distribution so as to ensure $\epsilon$-LDP guarantee. The other is that the crucial ingredient of PTT is a fundamental linear operation as will see later. It should be noted that the power of such a linear operation has been displayed in footnote 1.

From now on, let us elaborate on the development of PTT. To this end, we first need to introduce two elementary linear operations as follows.

\textbf{Definition 6} For an arbitrary element $x\in[-1,1]$, we can set up a couple of linear operations $L(x)$ and $R(x)$

\begin{equation}\label{eqa:MF-3-3-1}
L(x)=kx-a, \quad R(x)=kx+a, \quad k\neq 0,\quad a\neq 0.
\end{equation}
Then the resulted values $\widetilde{x}$ based on operations $L(x)$ and $R(x)$ compose of a range expressed as $\mathfrak{R}'(x,a)\triangleq[-k-a,k-a]\cup[-k+a,k+a]$. Without loss of generality, we only consider case of positive parameters $k$ and $a$. Other cases may be studied in a similar way, and thus are omitted.

Based on linear operations $L(x)$ and $R(x)$ in Def.6, we are able to construct a probability distribution from which the noise required to perturb attribute $A_{i}$ is drawn. For our purpose, we denote by $PDF(\widetilde{A_{i}}|A_{i})$ the probability density function of variable $\widetilde{A_{i}}$ given attribute $A_{i}$. And then, $PDF(\widetilde{A_{i}}|A_{i})$ is defined in the following form

\begin{equation}\label{eqa:MF-3-3-2}
PDF(\widetilde{A_{i}}|A_{i})=\left\{\begin{aligned}&f_{j}(\widetilde{A_{i}},A_{i}),\quad \text{if}\quad \widetilde{A_{i}}\in[L(A_{i}),R(A_{i})]\\
&\frac{p}{e^{\epsilon}}, \quad \text{if}\quad \widetilde{A_{i}}\in\mathfrak{R}(A_{i},a)-[L(A_{i}),R(A_{i})]
\end{aligned}
\right.,
\end{equation}
where $\mathfrak{R}(A_{i},a)$ is defined as interval $[-k-a,k+a]$, and bivariate functions $f_{j}(\widetilde{A_{i}},A_{i})$ is given by

$$f_{j}(\widetilde{A_{i}},A_{i})=p+g(\widetilde{A_{i}},A_{i}), \quad p\times\frac{1-e^{\epsilon}}{e^{\epsilon}} \leq g(\widetilde{A_{i}},A_{i})\leq 0.$$
In the meantime, we use $\mathfrak{J}$ to denote a set consisted of all indices $j$ that meet the requirements above.

To make our statements more concrete, let us give an example. For a given attribute $A_{i}$, the corresponding data owner $o_{i}$ can obtain a sanitized value $\widetilde{A_{i}}$ sampled from either $[L(A_{i}),R(A_{i})]$ with a predefined probability $q$ (which is derived exactly) or $\mathfrak{R}(A_{i},a)-[L(A_{i}),R(A_{i})]$ with the complementary probability $1-q$.

Using the results in Eq.(\ref{eqa:MF-3-3-2}), we propose PTT which is shown in Algorithm 2.

\begin{algorithm}
	\caption{Piecewise transformation technique for one-dimensional numeric data}
	\label{alg:Framwork}
	\begin{algorithmic}[1]

\Require
\quad attribute $A_{i}\in[-1,1]$;
		privacy budget $\epsilon$

\Ensure attribute $\widetilde{A_{i}}\in[-B,B]$
		\State Sample a random variable $\omega$ from $[0,1]$ uniformly
        \If {$\omega<q$}
					\State Sample $\widetilde{A_{i}}$ uniformly at random from $[L(A_{i}),R(A_{i})]$
\Else
\State Sample $\widetilde{A_{i}}$ uniformly at random from $\mathfrak{R}(A_{i},a)-[L(A_{i}),R(A_{i})]$
				\EndIf
\State
		\Return $\widetilde{A_{i}}$
	\end{algorithmic}
\end{algorithm}

It should be mentioned that a similar algorithm to PTT has been proposed by Wang \emph{et al} \cite{Wang-2019-ICDE} for addressing the issue that we are studying. On the other hand, there are significant differences between two algorithms as will be shown later. See the coming discussions for more detailed information. In essence, the algorithm in \cite{Wang-2019-ICDE} will be proved to be a special member in our PTT.

\textbf{Theorem 1} PTT obeys $\epsilon$-LDP.

\emph{Proof}. By Def.1, for an arbitrary pair of attributes $A_{i}$ and $A_{j}$ in $[-1,1]$ and for any sanitized value $\widetilde{A_{o}}$, we have

\begin{equation}\label{eqa:MF-3-3-3}
\max_{A_{i},A_{j}}\frac{PDF(\widetilde{A_{o}}|A_{i})}{PDF(\widetilde{A_{o}}|A_{j})}=e^{\epsilon}.
\end{equation}
This is complete. \qed

As specified in framework 1, one would like to let the designed $\epsilon$-LDP scheme ensure that the output results can be used as an unbiased estimation for a given raw data. Towards this end, some pre-processing and post-processing methods need to be adopted. In the following, we will take some concrete schemes as examples to show more details.  Now, let us divert our attention on determining the expectation associated with $A_{i}$ in order to ensure unbiasedness, which is what we focus on in this paper.

\subsubsection{Type-I piecewise transformation technique}

We are going to study a simple and important member in the proposed PTT, which is referred to as Type-I piecewise transformation technique (shorted as I-PTT). Specifically, I-PTT is obtained by setting $f_{j}(\widetilde{A_{i}},A_{i})\equiv p$ in Eq.(\ref{eqa:MF-3-3-2}).

\textbf{Theorem 2} I-PTT provides an unbiased estimation for arbitrary attribute $A_{i}$ if probability parameters $p$ and $q$ follow

\begin{equation}\label{eqa:MF-3-3-4}
p=\frac{1}{2ak}\frac{e^{\epsilon}}{e^{\epsilon}-1}, \qquad q=\frac{e^{\epsilon}}{k(e^{\epsilon}-1)}.
\end{equation}
Often, parameter $q$ is required to be no less than $1/2$. This is a convention in the field of DP. For example, such a requirement is found in Random Response \cite{Warner-1965}. Hence, we also only consider case of this kind. In principle, the other case, i.e., $q\leq1/2$, can be analyzed in a similar manner to the following one.

\emph{Proof}. From Algorithm 2, it is easy to obtain the following expression

\begin{equation}\label{eqa:MF-3-3-4-1}
\begin{aligned}\mathbb{E}_{\mathfrak{P}}[\widetilde{A_{i}}]&=\int_{-B}^{L(A_{i})}\frac{p}{e^{\epsilon}}xdx+\int_{L(A_{i})}^{R(A_{i})}pxdx+\int_{R(A_{i})}^{B}\frac{p}{e^{\epsilon}}xdx\\
&=\frac{p}{2e^{\epsilon}}(L^{2}(A_{i})-B^{2})+\frac{p}{2}(R^{2}(A_{i})-L^{2}(A_{i}))\\
&\quad+\frac{p}{2e^{\epsilon}}(B^{2}-R^{2}(A_{i}))\\
&=\frac{p}{2e^{\epsilon}}(L^{2}(A_{i})-R^{2}(A_{i}))+\frac{p}{2}(R^{2}(A_{i})-L^{2}(A_{i}))\\
&=\frac{p}{2}\frac{e^{\epsilon}-1}{e^{\epsilon}}(R^{2}(A_{i})-L^{2}(A_{i}))
\end{aligned},
\end{equation}
in which parameter $B$ is calculated to equal $k+a$.

To ensure unbiasedness, it is necessary to prove

\begin{equation}\label{eqa:MF-3-3-4-2}
\mathbb{E}_{\mathfrak{P}}[\widetilde{A_{i}}]=A_{i}.
\end{equation}
Solving for $p$ from the preceding equation yields

\begin{equation}\label{eqa:MF-3-3-4-3}
p=\frac{1}{2ak}\frac{e^{\epsilon}}{e^{\epsilon}-1}.
\end{equation}
Note that we have used Eq.(\ref{eqa:MF-3-3-1}). Furthermore, the predefined parameter $q$ is by definition given by

\begin{equation}\label{eqa:MF-3-3-4-4}
q=\int_{L(A_{i})}^{R(A_{i})}pdx=\frac{e^{\epsilon}}{k(e^{\epsilon}-1)}.
\end{equation}
This is complete. \qed

To make further progress, we want to measure the steadiness of all perturbed data $\widetilde{A_{i}}$ associated with attribute $A_{i}$ by mean of variance $\mathrm{Var}_{\mathfrak{P}}(\widetilde{A_{i}})$.

\textbf{Theorem 3} Given an arbitrary attribute $A_{i}$, after running I-PTT, the closed-form solution of variance $\mathrm{Var}_{\mathfrak{P}}(\widetilde{A_{i}})$ is given by
\begin{equation}\label{eqa:MF-3-3-5}
\mathrm{Var}_{\mathfrak{P}}(\widetilde{A_{i}})=(k-1)A^{2}_{i}+\frac{1}{3(\eta-1)}\left(\frac{\eta^{3}}{e^{\epsilon}-1}+1\right)a,
\end{equation}
where we consider I-PTT with unbiasedness and $\eta$ is equal to $B/a$. At the same time, it is worth mentioning that we only study PTTs of such type hereafter.

\emph{Proof}. By definition, we can write

\begin{equation}\label{eqa:MF-3-3-5-1}
\begin{aligned}\mathrm{Var}_{\mathfrak{P}}(\widetilde{A_{i}})&=\mathbb{E}_{\mathfrak{P}}[\widetilde{A_{i}}^{2}]-\mathbb{E}_{\mathfrak{P}}^{2}[\widetilde{A_{i}}]\\
&=\int_{-B}^{L(A_{i})}\frac{p}{e^{\epsilon}}x^{2}dx+\int_{L(A_{i})}^{R(A_{i})}px^{2}dx\\
&\quad+\int_{R(A_{i})}^{B}\frac{p}{e^{\epsilon}}x^{2}dx-A^{2}_{i}\\
&=\frac{p}{3e^{\epsilon}}(L^{3}(A_{i})+B^{3})+\frac{p}{3}(R^{3}(A_{i})-L^{3}(A_{i}))\\
&\quad+\frac{p}{3e^{\epsilon}}(B^{3}-R^{3}(A_{i}))-A^{2}_{i}\\
&=\frac{p}{3}\frac{e^{\epsilon}-1}{e^{\epsilon}}(R^{3}(A_{i})-L^{3}(A_{i}))+\frac{2p}{3e^{\epsilon}}B^{3}-A^{2}_{i}
\end{aligned},
\end{equation}
in which parameter $B$ is also equal to $k+a$. Also, we have used the unbiasedness of I-PTT.

Substituting Eq.(\ref{eqa:MF-3-3-1}) into Eq.(\ref{eqa:MF-3-3-5-1}) and implementing some fundamental arithmetics together yields

$$\mathrm{Var}_{\mathfrak{P}}(\widetilde{A_{i}})=(k-1)A^{2}_{i}+\frac{1}{3(\eta-1)}\left(\frac{\eta^{3}}{e^{\epsilon}-1}+1\right)a$$
as desired. This completes the proof of Theorem 3. \qed

From Eq.(\ref{eqa:MF-3-3-5}), we see that variance $\mathrm{Var}_{\mathfrak{P}}(\widetilde{A_{i}})$ is dependent on attribute $A_{i}$. This is similar to Duchi's scheme stated in Lemma 2. However, there is a considerable difference between I-PTT and Duchi's scheme with regard to variance. Meanwhile, for a given privacy budget $\epsilon$, the variance associated with Laplace mechanism is a constant, on the other hand, the presented I-PTT has a variance due to parameters $a$ and $\eta$. As a result, it is of great interest to find out the optimal I-PTT provided privacy budget $\epsilon$. In the meantime, it is also an attractive task to compare our scheme with both Laplace mechanism and Duchi's scheme by means of some measures such as variance. These issues are answered completely in the following discussions. Before beginning with our discussions, a previously published scheme needs to be revisited. In Ref.\cite{Wang-2019-ICDE}, the piecewise mechanism is studied analytically. In particular, this scheme turns out to be a special case of I-PTT in which parameters $k$ and $a$ in Eq.(\ref{eqa:MF-3-3-1}) are assumed to follow

\begin{equation}\label{eqa:MF-3-3-5-2}
k=\frac{2e^{\epsilon/2}}{e^{\epsilon/2}-1}, \quad a=\frac{2}{e^{\epsilon/2}-1}.
\end{equation}
Consequently, the solution of probability parameter $p$ in function $f_{j}(\widetilde{A_{i}},A_{i})$ is derived in the following form

\begin{equation}\label{eqa:MF-3-3-5-3}
p=\frac{e^{\epsilon}-e^{\epsilon/2}}{2(e^{\epsilon/2}+1)}.
\end{equation}
From Eq.(\ref{eqa:MF-3-3-5-3}), we can see that privacy budget $\epsilon$ can not take all values in interval $(0,+\infty)$, which is not pointed in \cite{Wang-2019-ICDE}. In a nutshell, the prior work in \cite{Wang-2019-ICDE} is covered by this work completely.

In the following, let us answer the first issue mentioned above, i.e., studying optimality of I-PTT itself. Fortunately, the results will mean that given a privacy budget $\epsilon$, one has the ability to select an optimal member from a family of I-PTTs when varying another parameter $a$.

\textbf{Theorem 4} For a given privacy budget $\epsilon$ and probability parameter $q$, there is a unique optimal I-PTT with respect to variance $\mathrm{Var}_{\mathfrak{P}}(\widetilde{A_{i}})$. In which case, parameter $a$ is given by

\begin{equation}\label{eqa:MF-3-3-6}
a=\frac{\frac{e^{\epsilon}}{q(e^{\epsilon}-1)}}{\sqrt[3]{e^{\epsilon}+\sqrt{(e^{\epsilon}+1)(e^{\epsilon}-1)}}+\sqrt[3]{e^{\epsilon}-\sqrt{(e^{\epsilon}+1)(e^{\epsilon}-1)}}}.
\end{equation}
It should be mentioned that the corresponding proof is deferred to show in Appendix A.

Based on Theorem 4, we can immediately come to Theorem 5 that ensures the accuracy guarantee of estimation $\widetilde{\mathfrak{A}}=\sum_{i=1}^{n}\widetilde{A_{i}}/n$. Quantity $\widetilde{\mathfrak{A}}$ is used by the data curator to estimate the mean value of all raw attributes $A_{i}$.

\textbf{Theorem 5} Suppose that $\widetilde{\mathfrak{A}}=\sum_{i=1}^{n}\widetilde{A_{i}}/n$ and $\mathfrak{A}=\sum_{i=1}^{n}A_{i}/n$. With at least $1-\beta$ probability, we gain

\begin{equation}\label{eqa:MF-3-3-7}
\Theta_{\mathfrak{P}}=|\widetilde{\mathfrak{A}}-\mathfrak{A}|=O\left(\frac{\sqrt{\log(1/\beta)}}{\epsilon\sqrt{n}}\right).
\end{equation}
As above, the corresponding proof is deferred to show in Appendix B.

From Eq.(\ref{eqa:MF-3-3-7}), we can see that as parameter $\epsilon\rightarrow0$, parameter $\gamma$ is of the same order of magnitude as  $\epsilon^{-1}n^{-1/2}$. It is worth mentioning that a similar conclusion also holds for both the typical Laplace mechanism and Duchi's scheme.
To some extent, we can say that the proposed I-PTT behaves a similar performance for ensuring estimation of mean of numerical data to these two well-known LDP schemes when only considering case of $\epsilon\rightarrow0$.

\subsubsection{Type-II piecewise transformation technique}

From now on, we propose another member in PTT and study some related properties. For brevity, we call it Type-II piecewise transformation technique (II-PTT for short). More precisely, function $f_{j}(\widetilde{A_{i}},A_{i})$ in Eq.(\ref{eqa:MF-3-3-2}) is now assumed as

$$f_{j}(\widetilde{A_{i}},A_{i})=p-\frac{p}{a}\frac{e^{\epsilon}-1}{e^{\epsilon}}\left|\widetilde{A_{i}}-(L(A_{i})+a)\right|.$$

\textbf{Theorem 6} II-PTT provides an unbiased estimation for arbitrary attribute $A_{i}$ if probability parameters $p$ and $q$ follow

\begin{equation}\label{eqa:MF-3-3-8-1}
p=\frac{1}{ak}\frac{e^{\epsilon}}{e^{\epsilon}-1}, \qquad q=\frac{e^{\epsilon}+1}{k(e^{\epsilon}-1)}.
\end{equation}
Meanwhile, the precise solution of variance $\mathrm{Var}_{\mathfrak{Q}}(\widetilde{A_{i}})$ is given by

\begin{equation}\label{eqa:MF-3-3-8-2}
\mathrm{Var}_{\mathfrak{Q}}(\widetilde{A_{i}})=(k-1)A^{2}_{i}+\frac{1}{6(\eta-1)}\left(\frac{4\eta^{3}}{e^{\epsilon}-1}+1\right)a,
\end{equation}
where $\eta$ is equal to $B/a$. The rigorous proof is deferred to shown in Appendix C.

Clearly, II-PTT has a variance dependent on attribute $A_{i}$, which is similar to observations from Eqs. (\ref{eqa:MF-3-2}) and (\ref{eqa:MF-3-3-5}). A natural problem is that which of them is optimal according to variance. In the following, we distinguish I-PTT and II-PTT. The comparisons with Duchi's scheme are deferred to discuss in the next subsection.

\textbf{Theorem 7} For a given pair of parameters $k$ and $a$, I-PTT is more optimal than II-PTT when privacy budget $\epsilon$ belongs to $(0,\ln3)$.

It should be mentioned that the mathematical proof is deferred to show in Appendix D.

As known, most previous work often focuses on the high privacy level particularly because of significantly theoretical interest and various practical applications \cite{Duchi-2018,Wang-2019-MOD}. The consequence above is saying that I-PTT is a good choice for the high privacy level by virtue of variance when given two candidates I-PTT and II-PTT.

By definition, a great variety of functions $f_{j}(\widetilde{A_{i}},A_{i})$ in Eq.(\ref{eqa:MF-3-3-2}) can be selected to create enough PTTs. As a consequence, we have a lot of potential schemes in the setting considered in this work. Here, we take I-PTT and II-PTT as illustrative examples to show the capability of PTT defined in Algorithm 2. Other cases can be analyzed in a similar manner as used above, and the  corresponding variances are derived either numerically or theoretically. Due to space limitation, we omit more details about discussions of this kind. One thing to note is that we are ware of the optimality of these candidate schemes particularly due to potential applications from various fields. Accordingly, this enables us to choose more reasonable schemes with which the issues given to us are effectively answered. Among of which, it is natural to ask which of these schemes is optimal with respect to some metrics, for example, variance. From Eq.(\ref{eqa:MF-3-3-9-1}), we can bring a conjecture as follows.

\textbf{Conjecture 1} At the high privacy level, I-PTT is the optimal with regard to variance.

\subsection{Comparisons between our schemes and previous ones}

In the above subsection, we focus mainly on comparisons between members in PTT. Now, the goal of this subsection is to distinguish ours schemes and previous ones, namely, the widely-used Laplace mechanism and Duchi's scheme. In view of various members in PTT, we only select I-PTT as an illustrative example to show more details. Other members may be analyzed in a similar way.

First of all, we need to introduce two notations upon variance in order to build metric for comparing our schemes with the previously published ones.

\textbf{Definition 7} Given two $\epsilon$-LDP schemes, say $\mathcal{M}_{1}$ and $\mathcal{M}_{2}$, the consistency noisy variance, denoted by $r(\epsilon,A_{i})$, is defined to as

\begin{equation}\label{eqa:MF-3-4-0-1}
r(\epsilon,A_{i})\triangleq\mathrm{Var}_{\mathcal{M}_{1}}(\widetilde{A_{i}})-\mathrm{Var}_{\mathcal{M}_{2}}(\widetilde{A_{i}}), \qquad \forall \quad A_{i}\in[-1,1]
\end{equation}
where $\mathrm{Var}_{\mathcal{M}_{1}}(\widetilde{A_{i}})$ and $\mathrm{Var}_{\mathcal{M}_{2}}(\widetilde{A_{j}})$ represent variance of schemes $\mathcal{M}_{1}$ and $\mathcal{M}_{2}$, respectively.

Often, the consistency noisy variance $r(\epsilon,A_{i})$ is a quite strong condition for comparing two $\epsilon$-LDP schemes with unbiasedness mean estimation. In the field of PPDA, one is often interested in another weaker version associated with variance, i.e., the worst-case noisy variance, and uses the weaker measure to valuate a given pair of $\epsilon$-LDP schemes. In the following, let us bring the concrete definition of the worst-case noisy variance from aspect of $\epsilon$-LDP scheme.

\textbf{Definition 8} Given two $\epsilon$-LDP schemes, say $\mathcal{M}_{1}$ and $\mathcal{M}_{2}$, the worst-case noisy variance, denoted by $s(\epsilon)$, is defined to as

\begin{equation}\label{eqa:MF-3-4-0-2}
s(\epsilon)\triangleq\max_{A_{i}}\mathrm{Var}_{\mathcal{M}_{1}}(\widetilde{A_{i}})-\max_{A_{j}}\mathrm{Var}_{\mathcal{M}_{2}}(\widetilde{A_{j}}) , \quad \forall A_{i}, A_{j}\in[-1,1]
\end{equation}
where $\mathrm{Var}_{\mathcal{M}_{1}}(\widetilde{A_{i}})$ and $\mathrm{Var}_{\mathcal{M}_{2}}(\widetilde{A_{j}})$ denote variance of schemes $\mathcal{M}_{1}$ and $\mathcal{M}_{2}$, respectively.

By definition, it is easy to understand that the worst-case noisy variance is independent of value of input $A_{i}$. This is why the worst-case noisy variance is a weaker measure than the consistency noisy variance. We are now ready to proceed to handle our tasks. First, I-PTT is compared with the well-known Duchi's scheme by means of the consistency noisy variance.

\textbf{Theorem 8} There are no more optimal I-PTTs than Duchi's scheme with respect to the consistent noisy variance.

Note that the rigorous proof is deferred to shown in Appendix E.

As mentioned above, the condition contained in Theorem 8 is quite strong. That is to say, the proposed I-PTT is considered optimal only if function $p_{1}(\epsilon)$ defined in Eq.(\ref{eqa:MF-3-4-1-2}) is completely equal to or smaller than zero for arbitrary $\epsilon(>0)$. Now, we switch gear to look into comparison between two schemes by virtue of the weaker metric in Def.8.

\textbf{Theorem 9} There are a great number of I-PTTs that are more optimal than Duchi's scheme with respect to the worst-case noisy variance.

\emph{Proof} First of all, according to Eqs.(\ref{eqa:MF-3-2}) and (\ref{eqa:MF-3-3-5}), the worst-case noisy variance between our I-PTT and Duchi's scheme is given by

\begin{equation}\label{eqa:MF-3-4-2-1}
\begin{aligned}s_{1}&(\epsilon)\triangleq\max_{A_{i}}\mathrm{Var}_{\mathfrak{P}}(\widetilde{A_{i}})-\max_{A_{j}}\mathrm{Var}_{\mathfrak{D}}(\widetilde{A_{j}})\\
&=(\eta-1)a-1+\frac{1}{3(\eta-1)}\left(\frac{\eta^{3}}{e^{\epsilon}-1}+1\right)a-\left(\frac{e^{\epsilon}+1}{e^{\epsilon}-1}\right)^{2}
\end{aligned}.
\end{equation}

Using a similar transformation technique as in proof of Theorem 8, we rewrite Eq.(\ref{eqa:MF-3-4-2-1}) as

\begin{equation}\label{eqa:MF-3-4-2-2}
s_{1}(t)=-4t^{2}-\left(4-\frac{a\eta^{3}}{3(\eta-1)}\right)t+\frac{a}{3(\eta-1)}+(\eta-1)a-2,
\end{equation}
where $t$ is equal to $1/(e^{\epsilon}-1)$. From here on out, it is sufficient to prove that maximum of function $s_{1}(t)$, i.e., $\max_{t}s_{1}(t)$, is strictly equal to or smaller than zero for the purpose of showing existence of optimal I-PTT. To this end, using Eq.(\ref{eqa:MF-3-3-4}), we can first define a function $p_{2}(t)$

\begin{equation}\label{eqa:MF-3-4-2-3}
\begin{aligned}p_{2}(t)&=\left(\frac{\eta^{3}}{3(\eta-1)^{2}}-4\right)t^{2}+\frac{1}{3(\eta-1)^{2}}-1\\
&\quad+\left(\frac{\eta^{3}}{3(\eta-1)^{2}}+\frac{1}{3(\eta-1)^{2}}-3\right)t
\end{aligned},
\end{equation}
and then prove that for a given parameter $\eta(>1)$, the next inequality holds

\begin{equation}\label{eqa:MF-3-4-2-4}
p_{2}(t)<\max_{t}s_{1}(t), \quad \forall \quad t\in(0,+\infty).
\end{equation}
From that we see that if there are functions $p_{2}(t)$ that turn out to be strictly smaller than zero for arbitrary $t$, then it will be very likely to find what we are interested in. Now, our task is to find available function $p_{2}(t)$. Towards this end, it suffices to solve for $\eta$ from either the coming inequality system

\begin{equation}\label{eqa:MF-3-4-2-5}
\left\{\begin{aligned}&\begin{aligned}f_{1}(\eta)&\triangleq\left(\frac{\eta^{3}}{3(\eta-1)^{2}}+\frac{1}{3(\eta-1)^{2}}-3\right)^{2}\\
&\quad-4\left(\frac{\eta^{3}}{3(\eta-1)^{2}}-4\right)\left(\frac{1}{3(\eta-1)^{2}}-1\right)\leq0
\end{aligned}\\
&f_{2}(\eta)\triangleq\eta^{3}-12\eta^{2}+24\eta-12<0
\end{aligned}
\right.,
\end{equation}
or
\begin{equation}\label{eqa:MF-3-4-2-6}
\left\{\begin{aligned}&\begin{aligned}f_{1}(\eta)&\triangleq\left(\frac{\eta^{3}}{3(\eta-1)^{2}}+\frac{1}{3(\eta-1)^{2}}-3\right)^{2}\\
&\quad-4\left(\frac{\eta^{3}}{3(\eta-1)^{2}}-4\right)\left(\frac{1}{3(\eta-1)^{2}}-1\right)>0\end{aligned}\\
&f_{2}(\eta)\triangleq\eta^{3}-12\eta^{2}+24\eta-12<0\\
&f_{3}(\eta)\triangleq\eta^{3}+3(\eta-1)^{2}-12\eta+13\leq0\\
&f_{4}(\eta)\triangleq1-3(\eta-1)^{2}\leq0
\end{aligned}
\right..
\end{equation}
In view of tedious calculations, we do not attempt to solve the above-mentioned inequality systems to derive candidate $\eta$. Yet, we illustrate range composed of parameter $\eta$ satisfying either Eq.(\ref{eqa:MF-3-4-2-5}) or Eq.(\ref{eqa:MF-3-4-2-6}) in Fig.2. See Fig.2 for more information. Here, in order to verify that there is indeed available $\eta$ such that our goal is achieved, we give a simple example as follows. It is easy to check that setting $\eta=19/10$ makes inequality system defined in Eq.(\ref{eqa:MF-3-4-2-5}) hold true. Next, according to Eq.(\ref{eqa:MF-3-3-4}), the exact solution of parameter $a$ may be assumed to obey the following value
$$a=\frac{10}{9}\times\frac{e^{\epsilon}+0.9}{e^{\epsilon}-1}$$
After that, Eq.(\ref{eqa:MF-3-4-2-2}) turns out to be always smaller than zero for an arbitrary parameter $\epsilon>0$. This suggests that we find optimal I-PTT. In this setting, two important parameters $p$ and $q$ are given by

\begin{figure}
\centering
  \includegraphics[height=6cm]{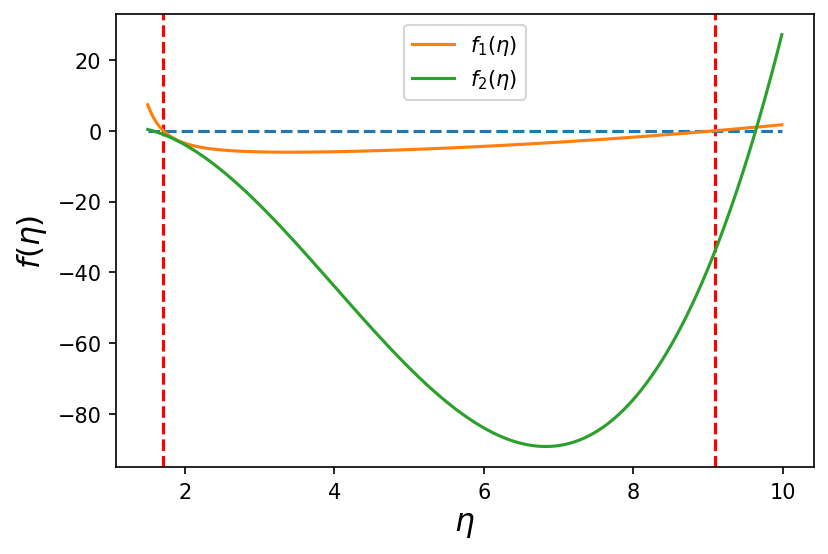}\\
{\small Fig.2. (Color online) The diagram of numerical solutions to parameter $\eta$ in Eq.(\ref{eqa:MF-3-4-2-5}). It is easy to see that all $\eta$ falling within the intermediate range indicated by two red dashed lines are subjected to constraint defined in Eq.(\ref{eqa:MF-3-4-2-5}). This suggests that there are sufficiently many potential $\eta$ such that a great variety of optimal schemes are created. In addition, the results means that there are no candidate parameter $\eta$ that makes requirement defined in Eq.(\ref{eqa:MF-3-4-2-6}) satisfied. Thus, the illustration associated with Eq.(\ref{eqa:MF-3-4-2-6}) is omitted.   }
\end{figure}

\begin{equation}\label{eqa:MF-3-4-2-8}
p=\frac{9}{20}\times\frac{e^{\epsilon}(e^{\epsilon}-1)}{(e^{\epsilon}+0.9)^{2}}, \quad q=\frac{e^{\epsilon}}{e^{\epsilon}+0.9}.
\end{equation}
Clearly, these two probability parameters are subject to requirement mentioned in Theorem 2.

Armed with the aforementioned results, we demonstrate that according to the worst-case noisy variance, one has the ability to construct a great number of optimal I-PTTs compared with Duchi's scheme. By far, we finish the proof of Theorem 9 completely. \qed

With Theorems 8 and 9, we indeed see that the consistency noisy variance is a stronger measure for distinguishing a pair of $\epsilon$-LDP schemes than the worst-case noisy variance. It is well known that the privacy guarantee of arbitrary $\epsilon$-LDP scheme is drastically degraded with increasing privacy budget $\epsilon$. At the same time, the intuition of building $\epsilon$-LDP scheme is to protect private information of individual data owner. Therefore, it is sufficient and interesting to only consider the high privacy level, i.e., $\epsilon\in(0,1]$, when analyzing $\epsilon$-LDP scheme.

The derived result is saying that in the worst-case noisy variance setting, the lower bound of variance of I-PTT, denoted by $\mathrm{Var}_{\mathfrak{P}}(\widetilde{A_{i}})^{Low}_{A_{i}=1}$, is bounded. More precisely, we have

\begin{equation}\label{eqa:MF-3-4-2-9}
\lim_{\epsilon\rightarrow 0}\mathrm{Var}_{\mathfrak{P}}(\widetilde{A_{i}})^{Low}_{A_{i}=1}\leq O\left(\epsilon^{-2}\right).
\end{equation}
Then, a natural problem is whether or not quantity $\mathrm{Var}_{\mathfrak{P}}(\widetilde{A_{i}})^{Low}_{A_{i}=1}$ is bounded below by a parameter. Fortunately, there is a positive answer to this problem.

\textbf{Theorem 10} As $\epsilon\rightarrow0$, the low bound corresponding to the worst-case variance of I-PTT, namely, $\mathrm{Var}_{\mathfrak{P}}(\widetilde{A_{i}})^{Low}_{A_{i}=1}$,  is given by

\begin{equation}\label{eqa:MF-3-4-3}
\mathrm{Var}_{\mathfrak{P}}(\widetilde{A_{i}})^{Low}_{A_{i}=1}=O\left(\epsilon^{-2}\right).
\end{equation}

\emph{Proof} Here, the worst-case variance of I-PTT is referred to as $\mathrm{Var}_{\mathfrak{P}}(\widetilde{A_{i}})_{A_{i}=1}$ for convenience. And then, the associated expression is rewritten as

\begin{equation}\label{eqa:MF-3-4-3-1}
\mathrm{Var}_{\mathfrak{P}}(\widetilde{A_{i}})_{A_{i}=1}=(\eta-1)a-1+\frac{1}{3(\eta-1)}\left(\frac{\eta^{3}}{e^{\epsilon}-1}+1\right)a,
\end{equation}
where we have used relationship $k=(\eta-1)a$.

From Eq.(\ref{eqa:MF-3-3-4}), we have

\begin{equation}\label{eqa:MF-3-4-3-2}
\frac{e^{\epsilon}}{(e^{\epsilon}-1)(\eta-1)}<a\leq\frac{2e^{\epsilon}}{(e^{\epsilon}-1)(\eta-1)}.
\end{equation}

Combining Eq.(\ref{eqa:MF-3-4-3-1}) with Eq.(\ref{eqa:MF-3-4-3-2}) yields

\begin{equation}\label{eqa:MF-3-4-3-3}
h_{1}(\epsilon,\eta)\triangleq\frac{1}{e^{\epsilon}-1}+\frac{e^{\epsilon}\eta^{3}}{3(e^{\epsilon}-1)^{2}(\eta-1)^{2}}+\frac{e^{\epsilon}}{3(e^{\epsilon}-1)(\eta-1)^{2}}.
\end{equation}
Obviously, inequality $\mathrm{Var}_{\mathfrak{P}}(\widetilde{A_{i}})_{A_{i}=1}>h_{1}(\epsilon,\eta)$ holds for arbitrarily available parameters $\epsilon$ and $\eta$.

Now, for our purpose, we need to introduce a function $g_{1}(\epsilon)$, which is defined in the following form

\begin{equation}\label{eqa:MF-3-4-3-4}
g_{1}(\epsilon)\triangleq\frac{\theta_{1}}{(e^{\epsilon}-1)^{2}}+\frac{\theta_{2}}{e^{\epsilon}-1},
\end{equation}
in which the exact solutions of parameters $\theta_{1}$ and $\theta_{2}$ are derived in the next discussions, respectively. After that, we can define another function based on functions $h_{1}(\epsilon,\eta)$ and $g_{1}(\epsilon)$, which is expressed as

\begin{equation}\label{eqa:MF-3-4-3-5}
h_{2}(\epsilon,\eta)\triangleq h_{1}(\epsilon,\eta)-g_{1}(\epsilon).
\end{equation}

Using a simple transformation $t=1/(e^{\epsilon}-1)$, $h_{2}(\epsilon,\eta)$ is recast as
\begin{equation}\label{eqa:MF-3-4-3-6}
h_{2}(t,\eta)=\frac{1}{3(\eta-1)^{2}}\left\{\begin{aligned}&(\eta^{3}-3\theta_{1}(\eta-1)^{2})t^{2}+1\\
&+[\eta^{3}-3(\theta_{2}-1)(\eta-1)^{2}+1]t\end{aligned}\right\}.
\end{equation}
For the sake of brevity, we introduce two functions $\Psi_{i}(\eta)$ as below

\begin{subequations}
\label{eq:whole}
\begin{eqnarray}
\Psi_{1}(\eta)=\eta^{3}-3\theta_{1}(\eta-1)^{2},\label{subeq:MF-3-4-3-7-1}
\end{eqnarray}
\begin{equation}
\Psi_{2}(\eta)=\eta^{3}-3(\theta_{2}-1)(\eta-1)^{2}+1.\label{subeq:MF-3-4-3-7-2}
\end{equation}
\end{subequations}
And then, it is easy to prove that for parameter $\eta>1$, functions $\Psi_{i}(\eta)$ are no less than zero, respectively, if two parameters $\theta_{1}$ and $\theta_{2}$ obey the following expressions

\begin{equation}\label{eqa:MF-3-4-3-8}
\theta_{1}=\frac{9}{4}, \quad \theta_{2}=1+\frac{1}{3}\times\frac{\left(1+\sqrt[3]{2+\sqrt{3}}+\sqrt[3]{2-\sqrt{3}}\right)^{3}+1}{\left(\sqrt[3]{2+\sqrt{3}}+\sqrt[3]{2-\sqrt{3}}\right)^{2}}.
\end{equation}

Substituting results in Eq.(\ref{eqa:MF-3-4-3-8}) into Eq.(\ref{eqa:MF-3-4-3-6}), we have

\begin{equation}\label{eqa:MF-3-4-3-9}
h_{2}(t,\eta)=\frac{1}{3(\eta-1)^{2}}(\Psi_{1}(\eta)t^{2}+\Psi_{2}(\eta)t+1)>0
\end{equation}
where parameter $\epsilon$ in $(0,+\infty)$ and $\eta>1$. This further implies

\begin{equation}\label{eqa:MF-3-4-3-10}
\mathrm{Var}_{\mathfrak{P}}(\widetilde{A_{i}})_{A_{i}=1}> \frac{\theta_{1}}{(e^{\epsilon}-1)^{2}}+\frac{\theta_{2}}{e^{\epsilon}-1}.
\end{equation}
With the help of relationship $e^{\epsilon}-1\simeq\epsilon$ as $\epsilon\rightarrow0$, the correctness of Eq.(\ref{eqa:MF-3-4-3}) is consolidated based on Eq.(\ref{eqa:MF-3-4-3-10}). It should be mentioned that Eq.(\ref{eqa:MF-3-4-2-2}) has been used. Note also that the derived lower bound is tight. By far, we complete the proof of Theorem 10. \qed

More generally, we are not necessary concerned with the worst-case variance. That is to say, we may study case of varying $A_{i}$. In this sense, quantity $\mathrm{Var}_{\mathfrak{P}}(\widetilde{A_{i}})$ is bounded below as privacy budget $\epsilon$ approaches zero.

\textbf{Corollary 1} For each $A_{i}\in[-1,1]$, the lower bound of corresponding variance of I-PTT, denoted by $\mathrm{Var}_{\mathfrak{P}}(\widetilde{A_{i}})^{Low}_{A_{i}}$, is expressed as

\begin{equation}\label{eqa:MF-3-4-4}
\mathrm{Var}_{\mathfrak{P}}(\widetilde{A_{i}})^{Low}_{A_{i}}=O\left(\epsilon^{-2}\right),
\end{equation}
in which condition $\epsilon\rightarrow0$ has been adopted.

\emph{Proof} The correctness of Eq.(\ref{eqa:MF-3-4-4}) is verified in a similar fashion as used in proof of Theorem 10, and we thus omit it here.

Furthermore, with results in Eqs.(\ref{eqa:MF-3-1}), (\ref{eqa:MF-3-2}) and (\ref{eqa:MF-3-4-4}), a general relationship is obtained as follows: As $\epsilon\rightarrow0$,

\begin{equation}\label{eqa:MF-3-4-5} \mathrm{Var}_{\mathfrak{P}}(\widetilde{A_{i}})^{Low}_{A_{i}}\leq\mathrm{Var}_{\mathfrak{D}}(\widetilde{A_{i}})^{Low}_{A_{i}}\leq\mathrm{Var}_{\mathfrak{L}}(\widetilde{A_{i}})^{Low}_{A_{i}}=O\left(\epsilon^{-2}\right),
\end{equation}
where the meanings of notations $\mathrm{Var}_{\mathfrak{D}}(\widetilde{A_{i}})^{Low}_{A_{i}}$ and $\mathrm{Var}_{\mathfrak{L}}(\widetilde{A_{i}})^{Low}_{A_{i}}$ are clarified in the same manner as used to define $\mathrm{Var}_{\mathfrak{P}}(\widetilde{A_{i}})^{Low}_{A_{i}}$ in Corollary 1.

Now, we have reason to believe that the coming conjecture is true.

\textbf{Conjecture 2} There are optimal I-PTTs compared with the classic Laplace mechanism when considering the worst-case variance.

Although we do not provide a rigorous proof to Conjecture 2 at present, its rational is easy to prove given privacy budget $\epsilon$ in $(0,1]$. As known, it is of great interest to pay more attention on the high privacy level, which is a common sense agreement in the literature of LDP. Therefore, we omit discussion associated with the wide spectrum of parameter $\epsilon$. In other words, the complete proof to Conjecture 2 is left as our next move. In the meantime, this is left for interested reader as an exercise.

So far, we have finished the development of PTT and the analysis of two representatives, i.e., I-PTT and II-PTT. To close this section, we briefly summarize main results as follows. In this section, we propose a novel PTT for analyzing numerical attribute in an $\epsilon$-LDP manner. After that, two specific schemes I-PTT and II-PTT are proved to ensure an unbiased estimation for mean value associated with arbitrary attribute. At the same time, we derive the closed-form solution of noisy variance. Next, we show that for a given couple of privacy budget $\epsilon$ and probability parameter $q$, there is a unique optimal I-PTT that makes noisy variance minimal. At last, we analyze in detail differences between the presented I-PTT and two previously published schemes, i.e., Laplace mechanism and Duchi's scheme, according to some standard indices. Interested readers can refer Theorems 7-10 for more information. It is worth mentioning that one has the ability to find out sufficiently many I-PTTs, which are more optimal than these two prior schemes, by selecting available value for parameter $\eta$ when considering the high privacy level. In addition, we need to point out that in view of the results obtained above, a significant difference between our I-PTT and piecewise mechanism attributed to Wang \emph{et al} \cite{Wang-2019-ICDE} can be observed. More specifically, as reported in Ref.\cite{Wang-2019-ICDE}, the published piecewise mechanism is not always an optimal scheme compared with Duchi's scheme under the worst-case noisy variance. As shown in Theorem 9, however, it is possible to put forward many I-PTTs that are consistently optimal than Duchi's scheme under the same setting by selecting value of parameter $\eta$ carefully.  Lastly, in some senses, the proposed PTT actually reflects a fact that for an input $A_{i}$, the probability of observing each of a pair of possible distinct outputs $A_{j}$ and $A_{l}$ is always identical if they are a couple of symmetric points in terms of image $kA_{i}$ of $A_{i}$. From a methodological point of view, such a feature behind our PTT is similar to that hid in the classic Laplace mechanism.

As an immediate extension of PTT to multidimensional numerical data (without loss of generality, suppose that dimension is equal to $d$), we reach the following algorithm.

\begin{algorithm}
	\caption{PTT for multidimensional numerical data}
	\label{alg:Framwork}
	\begin{algorithmic}[1]

\Require
\quad tuple $A_{i}\in[-1,1]^{d}$;
		privacy budget $\epsilon$

\Ensure
 tuple $A^{\dagger}_{i}\in[-Bd,Bd]^{d}$
		\State Let $A^{\dagger}_{i}=\langle0,0,\dots,0\rangle$

\State Sample one value $j$ uniformly from $\{1,2,\dots,d\}$
\For {the sampled value $j$}
   \State Feed $A_{i}[j]$ and $\epsilon$ as input to PTT, and obtain a noisy value $x_{i,j}$
   \State $A^{\dagger}_{i}[j]=d\times x_{i,j}$
\EndFor
\State
		\Return $A^{\dagger}_{i}$
	\end{algorithmic}
\end{algorithm}

\textbf{Theorem 11} Suppose that $\widetilde{\mathfrak{A}}_{j}=\sum_{i=1}^{n}\widetilde{A_{i}[j]}/n$ and $\mathfrak{A}_{j}=\sum_{i=1}^{n}A_{i}[j]/n$. With at least $1-\beta$ probability, we gain

\begin{equation}\label{eqa:MF-3-4-8}
\max_{j\in\{1,\dots,d\}}\Theta_{\mathfrak{P}_{j}}=|\widetilde{\mathfrak{A}}_{j}-\mathfrak{A}_{j}|=O\left(\frac{\sqrt{d\log(d/\beta)}}{\epsilon\sqrt{n}}\right).
\end{equation}
Note that we omit the mathematical proof mainly because the correctness of Eq.(\ref{eqa:MF-3-4-8}) is validated in a similar manner as used in Appendix B.

In principle, the proposed PTT is based on a pair of simple transformation techniques that are shown in Def.6. It should be mentioned that there is a fundamental feature in Eq.(\ref{eqa:MF-3-3-1}). Precisely, for an arbitrary variable $x$, quantity $|R(x)-L(x)|$ is invariable. As an immediate result, our PTT is in fact established by virtue of piecewise transformation technique of constant length. Naturally, one can ask whether there are piecewise transformation technique of tuning length for constructing $\epsilon$-LDP with an unbiasedness mean estimator for arbitrary input. In which case, quantity $|R(x)-L(x)|$ is regarded as functions of variable $x$. We leave this open question as our next move.

Come what may, this kind of piecewise transformation techniques can be always studied in a fairly succinct way. It is a convention for people to design as simple schemes as possible in order to address a given issue. Clearly, the design principle behind our schemes is well consistent with the aforementioned intuition. In some sense, we come up with PTT by adhering to the famous Occam's Razor principle. In a word, taking into account the abovementioned merits, we would like to see that the proposed PTT can be selected as a fundamental building-block to build up various $\epsilon$-LDP protocols and related variants for numerical data analysis in the future.

\section{Related work}

Differential privacy, as a de facto standard for analyzing data while providing privacy-preserving guarantee, has received more attention from various kinds of fields \cite{Dwork-2006,Dwork-2006-1,Awan-2020,Li-2019}. In recent years, the corresponding local version, i.e., locally differential privacy, has merged as a tool that has been proven useful for analyzing data \cite{Imola-2021}-\cite{Duchi-2018}. As opposed to DP, there is no the so-called curator in LDP. The latter assumption is frequently encountered in many real-world scenarios \cite{Imola-2021}-\cite{Pastore-2021}. Therefore, in this study, we are concerned with a well-study analysis task, i.e., mean estimation of numerical data, in the LDP setting.

Roughly speaking, there are two classes of ways to address the issue mentioned above in the literature of LDP \cite{Dwork-2006,Duchi-2018}. The first class is built based on the following thought: converting an arbitrary numerical value into some of a group of discrete values with a predefined probability. The most famous among them is the scheme attributed to Duchi \emph{et al} \cite{Duchi-2018}. This scheme has been used as an ingredient to design various LDP-protocols, such as Harmony-mean \cite{Nguyen-2016}, PrivKVM \cite{Ye-2019} and PCKV \cite{Gu-2020}. The other class is upon another thought: converting an arbitrary numerical value into some of a set of numerical values with a predefined probability density function. Among them, the best-known is the classic Laplace mechanism due to Dwork \cite{Dwork-2006}. Along the same research line, many similar protocols have been proposed, for instance, a complicated variant of Laplace mechanism, staircase mechanism \cite{Geng-2015} and piecewise mechanism \cite{Wang-2019-ICDE}. Our work falls into the second class. As mentioned in the preceding section, the piecewise mechanism in \cite{Wang-2019-ICDE} turns out to be a specific member of our PTT. It should be mentioned that this work aims to establish fundamental LDP-scheme, which can be able to be adopted as building-block to construct more complicated LDP-protocols suitable for various scenarios.

Additionally, LDP has been deployed into a great variety of practical applications including Apple's iOS, macOS and Safari \cite{Apple-2016}, Microsoft Windows 10 \cite{Ding-2017}, Google Chrome software systems \cite{Erlingsson-2014}, and Alibaba \cite{Wang-2019-MOD}. At the same time, it has been shown that LDP is also suitable for tackling many other analysis tasks, for example frequency/histogram estimation on categorical data \cite{Wang-2017-USNIX} and discovering heavy hitter \cite{Qin-2016}, frequent item/itemset mining of itemset data \cite{Wang-2018-SP}, generating synthetic social graphs \cite{Wei-2020} and counting subgraphs on graph \cite{Imola-2021}.

\section{Conclusion}

In summary, we are particularly concerned with privacy preserving data analysis under local differential privacy. More precisely, two commonly-studied topics in PPDA, namely, mean value and frequency estimation on numerical data, are analyzed in more detail. In this work, we put forward a new transformation technique, called PTT, for obtaining an unbiased estimator of mean value while satisfying the rigorous requirement defined by LDP. More importantly, we build up a principled framework for PTT, based on which the proposed transformation technique is discussed systematically. The results show that (1) there exist a great number of members in PTTs that are asymptotically optimal when utilized to capture an unbiased estimator for mean of numerical data, and (2) for a given privacy budget, one is able to design optimal PTT that reaches the theoretical low bound with respect to noisy variance. Next, we distinguish between the proposed PTT and two famous and fundamental protocols, i.e., Laplace mechanism and Duchi's scheme, in the realm of LDP. The results suggest that (1) when analyzing numerical data under LDP, there do not exist optimal PTTs compared to Duchi's scheme in terms of the consistency noisy variance, (2) however, one has the ability to find too many PTTs that are consistently more optimal than the latter with regard to the worst-case noisy variance. Also, if we are limited to discussions at the high privacy level, then we can definitely to present optimal PTTs contrasted with the classic Laplace mechanism. At last, we obtain that for I-PTT, the correspondingly theoretical low bound of noisy variance follows $O(\epsilon^{-2})$ as privacy budget $\epsilon$ goes to zero.

\section*{Acknowledgment}

The research was supported by the National Key Research and Development Plan under grant 2020YFB1805400 and the National Natural Science Foundation of China under grant No. 62072010.

\section*{Appendix}

The section aims to provide many auxiliary information including proof to theorem for reader to well understand contents in main text.

\subsection*{Appendix A: Proof of Theorem 4 }

\setcounter{equation}{0}
\renewcommand\theequation{A.\arabic{equation}}

\emph{Proof} It suffices to derive minimum of variance $\mathrm{Var}_{\mathfrak{P}}(\widetilde{A_{i}})$ as a function of variable $\eta(>1)$. Using Eq.(\ref{eqa:MF-3-3-4}), Eq.(\ref{eqa:MF-3-3-5}) is rewritten as

\begin{equation}\label{eqa:MF-3-3-6-1}
\mathrm{Var}_{\mathfrak{P}}(\widetilde{A_{i}})=(k-1)A^{2}_{i}+\frac{k}{3(\eta-1)^{2}}\left(\frac{\eta^{3}}{e^{\epsilon}-1}+1\right).
\end{equation}

Performing derivative on both-hand sides of Eq.(\ref{eqa:MF-3-3-6-1}) with respect to $\eta$ leads to

\begin{equation}\label{eqa:MF-3-3-6-2}
\frac{d}{d\eta}\mathrm{Var}_{\mathfrak{P}}(\widetilde{A_{i}})=\frac{k}{3(e^{\epsilon}-1)(\eta-1)^{3}}\left[\eta^{3}-3\eta^{2}-2(e^{\epsilon}-1)\right].
\end{equation}

Now, we define a new function $f(\eta)$, i.e., $f(\eta)\triangleq\eta^{3}-3\eta^{2}-2(e^{\epsilon}-1)$. And then, taking derivative on function $f(\eta)$ with respect to $\eta$ yields

\begin{equation}\label{eqa:MF-3-3-6-3}
\frac{d}{d\eta}f(\eta)=3\eta(\eta-2).
\end{equation}
From that we can easily see that $\mathrm{Var}_{\mathfrak{P}}(\widetilde{A_{i}})$ is a monotone decreasing function for $\eta\in(1,\eta_{0}]$ and a monotone increasing function for $\eta\in[\eta_{0},+\infty)$ where $\eta_{0}$ is a non-negative root of function $f(\eta)$. This suggests that there is a unique optimal I-PTT according to variance $\mathrm{Var}_{\mathfrak{P}}(\widetilde{A_{i}})$. After performing some fundamental computations, the exact solution of quantity $\eta_{0}$ is given by

\begin{equation}\label{eqa:MF-3-3-6-4}
\eta_{0}=1+\sqrt[3]{e^{\epsilon}+\sqrt{(e^{\epsilon}+1)(e^{\epsilon}-1)}}+\sqrt[3]{e^{\epsilon}-\sqrt{(e^{\epsilon}+1)(e^{\epsilon}-1)}}.
\end{equation}

After some algebra, we can derive the precise solution of $a$ as shown in Eq.(\ref{eqa:MF-3-3-6}). Note that Fig.3 shows some numerical results related to variance $\mathrm{Var}_{\mathfrak{P}}(\widetilde{A_{i}})$ in view of distinct values of parameters $q$ and $\epsilon$. Thus far, we complete the proof of Theorem 4. \qed

\begin{figure*}
\centering
  \centering
\subfigure[$\epsilon=\frac{1}{2},\quad q=\frac{1}{2}$]{
\begin{minipage}[t]{0.23\linewidth}
\centering
\includegraphics[width=4cm]{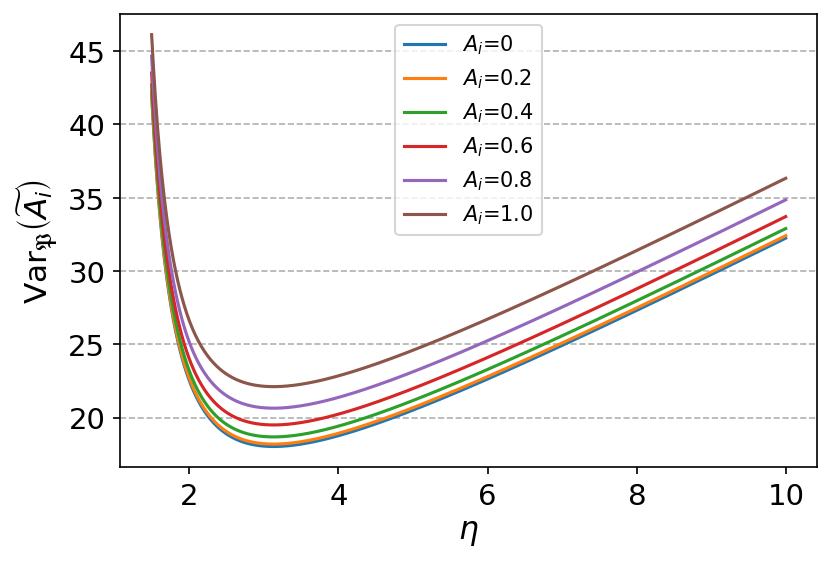}
\end{minipage}
}%
\subfigure[$\epsilon=\frac{1}{2},\quad q=\frac{3}{4}$]{
\begin{minipage}[t]{0.23\linewidth}
\centering
\includegraphics[width=4cm]{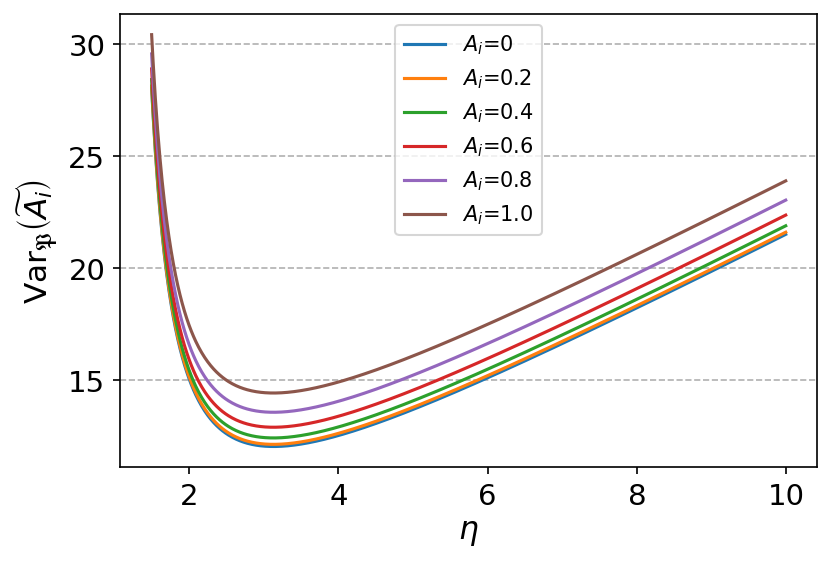}
\end{minipage}
}%
\subfigure[$\epsilon=1,\quad q=\frac{1}{2}$]{
\begin{minipage}[t]{0.23\linewidth}
\centering
\includegraphics[width=4cm]{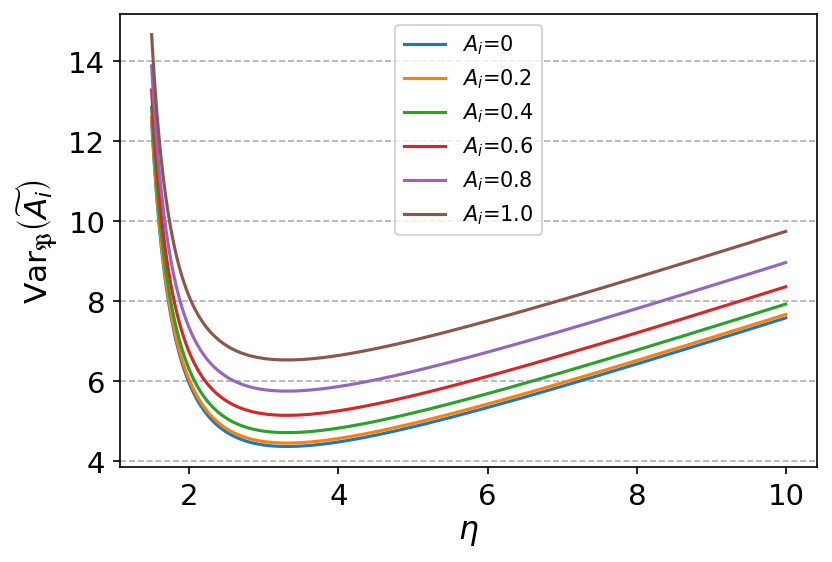}
\end{minipage}
}%
\subfigure[$\epsilon=1,\quad q=\frac{3}{4}$]{
\begin{minipage}[t]{0.23\linewidth}
\centering
\includegraphics[width=4cm]{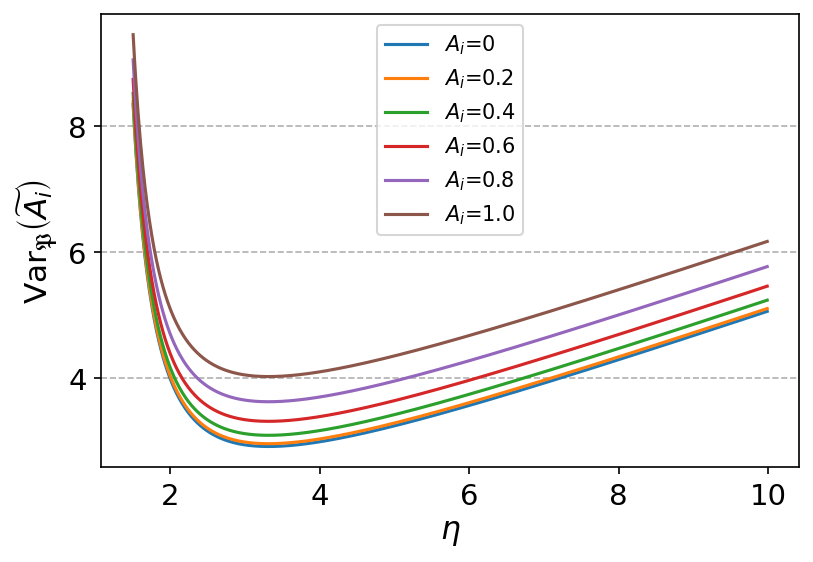}
\end{minipage}
}\\
{\small Fig.2. (Color online) The diagram of numerical solutions to parameter $\eta$ in Eq.(\ref{eqa:MF-3-4-2-5}). It is easy to see that all $\eta$ falling within the intermediate range indicated by two red dashed lines are subjected to constraint defined in Eq.(\ref{eqa:MF-3-4-2-5}). This suggests that there are sufficiently many potential $\eta$ such that a great variety of optimal schemes are created. In addition, the results means that there are no candidate parameter $\eta$ that makes requirement defined in Eq.(\ref{eqa:MF-3-4-2-6}) satisfied. Thus, the illustration associated with Eq.(\ref{eqa:MF-3-4-2-6}) is omitted.   }
\end{figure*}

\subsection*{Appendix B: Proof of Theorem 5 }

\setcounter{equation}{0}
\renewcommand\theequation{B.\arabic{equation}}

\emph{Proof} For a given attribute $A_{i}$, Theorem 1 means that difference between $\widetilde{\mathfrak{A}}$ and $\mathfrak{A}$, i.e., $\Theta_{\mathfrak{P}}=|\widetilde{\mathfrak{A}}-\mathfrak{A}|$, has zero mean. Meanwhile, Algorithm 2 indicates that quantity $\Theta_{\mathfrak{P}}$ is bounded by $k+a+1$. Based on this, using Bernstein's inequality, we have

\begin{equation}\label{eqa:MF-3-3-7-1}
\begin{aligned}\mathrm{Pr}&[|\widetilde{\mathfrak{A}}-\mathfrak{A}|>\gamma]=\mathrm{Pr}\left[\sum_{i=1}^{n}|\widetilde{A_{i}}-A_{i}|>n\gamma\right]\\
&\leq2\exp\left[-(n\gamma)^{2}\times\left(2\sum_{i=1}^{n}\mathrm{Var}_{\mathfrak{P}}(\widetilde{A_{i}})+\frac{2}{3}nm\gamma\right)^{-1}\right]
\end{aligned},
\end{equation}
in which parameter $m$ is required to obey

\begin{equation}\label{eqa:MF-3-3-7-2}
\mathrm{Pr}[|\widetilde{A_{i}}-A_{i}|>m]=0, \qquad \forall \quad A_{i}\in[-1,1].
\end{equation}
For our purpose, parameter $m$ may be assumed to equal $k+a+1$. Then, according to Eqs.(\ref{eqa:MF-3-3-5}) and (\ref{eqa:MF-3-3-7-1}), we can derive the asymptotic solution of parameter $\gamma$ in a mathematically rigorous manner, which is expressed as below

\begin{equation}\label{eqa:MF-3-3-7-3}
\gamma=O\left(\frac{\sqrt{\log(1/\beta)}}{\epsilon\sqrt{n}}\right).
\end{equation}
Notice that we have made use of condition $\epsilon\rightarrow 0$. Namely, we consider the high privacy level. Note also that the bound in Eq.(\ref{eqa:MF-3-3-7}) is asymptotically optimal \cite{Duchi-2018}. This completes the proof of Theorem 5. \qed

\subsection*{Appendix C: Proof of Theorem 6 }

\setcounter{equation}{0}
\renewcommand\theequation{C.\arabic{equation}}

\emph{Proof} In a similar way as used to prove Theorem 2, we obtain

\begin{equation}\label{eqa:MF-3-3-8-3}
\begin{aligned}\mathbb{E}_{\mathfrak{Q}}[\widetilde{A_{i}}]&=\int_{-B}^{L(t_{i})}\frac{p}{e^{\epsilon}}xdx++\int_{R(t_{i})}^{B}\frac{p}{e^{\epsilon}}xdx\\
&\quad\int_{L(A_{i})}^{L(A_{i})+a}\left[p-\frac{p}{a}\frac{e^{\epsilon}-1}{e^{\epsilon}}(L(A_{i})+a-x)\right]xdx\\
&\quad+\int_{L(A_{i})+a}^{R(t_{i})}\left[p-\frac{p}{a}\frac{e^{\epsilon}-1}{e^{\epsilon}}(x-L(A_{i})-a)\right]xdx\\
&=\frac{p}{2}\frac{e^{\epsilon}-1}{e^{\epsilon}}(R^{2}(A_{i})-L^{2}(A_{i}))-apk\frac{e^{\epsilon}-1}{e^{\epsilon}}A_{i}
\end{aligned}.
\end{equation}

By definition, after setting the expression on the last line in Eq.(\ref{eqa:MF-3-3-8-3}) to $A_{i}$, parameter $p$ is calculated to follow

\begin{equation}\label{eqa:MF-3-3-8-4}
p=\frac{1}{ak}\frac{e^{\epsilon}}{e^{\epsilon}-1}.
\end{equation}

Furthermore, parameter $q$ is expressed as
\begin{equation}\label{eqa:MF-3-3-8-5}
\begin{aligned}q&=\int_{L(A_{i})}^{L(A_{i})+a}\left[p-\frac{p}{a}\frac{e^{\epsilon}-1}{e^{\epsilon}}(L(A_{i})+a-x)\right]dx\\
&\quad+\int_{L(A_{i})+a}^{R(A_{i})}\left[p-\frac{p}{a}\frac{e^{\epsilon}-1}{e^{\epsilon}}(x-L(A_{i})-a)\right]dx\\
&=\frac{e^{\epsilon}+1}{k(e^{\epsilon}-1)}
\end{aligned}.
\end{equation}
Here, we have used Eq.(\ref{eqa:MF-3-3-8-4}).

Next, by definition, the exact expression of variance $\mathrm{Var}_{\mathfrak{Q}}(\widetilde{A_{i}})$ is written as

\begin{equation}\label{eqa:MF-3-3-8-6}
\begin{aligned}\mathrm{Var}_{\mathfrak{Q}}(\widetilde{A_{i}})&=\int_{-B}^{L(A_{i})}\frac{p}{e^{\epsilon}}x^{2}dx+\int_{R(A_{i})}^{B}\frac{p}{e^{\epsilon}}x^{2}dx-A_{i}^{2}\\
&\quad\int_{L(A_{i})}^{L(A_{i})+a}\left[p-\frac{p}{a}\frac{e^{\epsilon}-1}{e^{\epsilon}}(L(A_{i})+a-x)\right]x^{2}dx\\
&\quad+\int_{L(A_{i})+a}^{R(A_{i})}\left[p-\frac{p}{a}\frac{e^{\epsilon}-1}{e^{\epsilon}}(x-L(A_{i})-a)\right]x^{2}dx\\
&=(k-1)A^{2}_{i}+\frac{2B^{3}}{3a(B-a)(e^{\epsilon}-1)}+\frac{a^{2}}{6(B-a)}
\end{aligned}.
\end{equation}
Plugging relationship $B=\eta a$ into Eq.(\ref{eqa:MF-3-3-8-6}) yields the same result as in Eq.(\ref{eqa:MF-3-3-8-2}). This completes the proof of Theorem 6. \qed

\subsection*{Appendix D: Proof of Theorem 7}

\setcounter{equation}{0}
\renewcommand\theequation{D.\arabic{equation}}

\emph{Proof} It is sufficient to determine sign of the following expression

\begin{equation}\label{eqa:MF-3-3-9-1}
\begin{aligned}\mathrm{Var}_{\mathfrak{Q}}(\widetilde{A_{i}})-\mathrm{Var}_{\mathfrak{P}}(\widetilde{A_{i}})&=\frac{1}{6(\eta-1)}\left(\frac{4\eta^{3}}{e^{\epsilon}-1}+1\right)a\\
&\quad-\frac{1}{3(\eta-1)}\left(\frac{\eta^{3}}{e^{\epsilon}-1}+1\right)a\\
&=\frac{1}{6(\eta-1)}\left(\frac{2\eta^{3}}{e^{\epsilon}-1}-1\right)a
\end{aligned}.
\end{equation}

From Eq.(\ref{eqa:MF-3-3-1}), we see $\eta>1$. Because of this, it is easy to prove that the result on the second line in Eq.(\ref{eqa:MF-3-3-9-1}) is strictly larger than zero when parameter $\epsilon$ falls into interval $(0,\ln3)$. We complete the proof of Theorem 7.\qed

\subsection*{Appendix D: Proof of Theorem 8}

\setcounter{equation}{0}
\renewcommand\theequation{E.\arabic{equation}}

\emph{Proof} By definition, we need to first define

\begin{equation}\label{eqa:MF-3-4-1-1}
\begin{aligned}r_{1}(\epsilon,A_{i})&\triangleq\mathrm{Var}_{\mathfrak{P}}(\widetilde{A_{i}})-\mathrm{Var}_{\mathfrak{D}}(\widetilde{A_{i}})\\
&=(\eta-1)aA^{2}_{i}+\frac{1}{3(\eta-1)}\left(\frac{\eta^{3}}{e^{\epsilon}-1}+1\right)a\\
&\quad-\left(\frac{e^{\epsilon}+1}{e^{\epsilon}-1}\right)^{2}
\end{aligned}.
\end{equation}

For our purpose, it is sufficient to consider

\begin{equation}\label{eqa:MF-3-4-1-2}
\begin{aligned}p_{1}(\epsilon)&=\max_{\mbox{\tiny$A_{i}$}} r_{1}(\epsilon,A_{i})\\
&=(\eta-1)a+\frac{1}{3(\eta-1)}\left(\frac{\eta^{3}}{e^{\epsilon}-1}+1\right)a-\left(\frac{e^{\epsilon}+1}{e^{\epsilon}-1}\right)^{2}.
\end{aligned}
\end{equation}

Setting $t=\frac{1}{e^{\epsilon}-1}$, Eq.(\ref{eqa:MF-3-4-1-2}) is rearranged as

\begin{equation}\label{eqa:MF-3-4-1-3}
p_{1}(t)=-4t^{2}-\left(4-\frac{a\eta^{3}}{3(\eta-1)}\right)t+\frac{a}{3(\eta-1)}+(\eta-1)a-1.
\end{equation}
Obviously, we see

\begin{equation}\label{eqa:MF-3-4-1-4}
\begin{aligned}p_{1}\left(-\frac{4-\frac{a\eta^{3}}{3(\eta-1)}}{8}\right)&=\left(\frac{a\eta^{3}}{12(\eta-1)}-1\right)^{2}\\
&\quad+\frac{a}{3(\eta-1)}+(\eta-1)a-1>0
\end{aligned}.
\end{equation}
Hence, if we want to find optimal I-PTT, it is necessary to require that the following inequalities hold true

\begin{equation}\label{eqa:MF-3-4-1-5}
\left\{\begin{aligned}&4-\frac{a\eta^{3}}{3(\eta-1)}\geq0, \quad \forall \quad \eta>1\\
&\lim_{t\rightarrow 0}p(t)\leq0
\end{aligned}
\right..
\end{equation}
From Eq.(\ref{eqa:MF-3-3-4}), the left-hand side on the first line of Eq.(\ref{eqa:MF-3-4-1-5}) is bounded below by the coming expression

\begin{equation}\label{eqa:MF-3-4-1-6}
I_{1}(\eta)\triangleq4-\frac{2e^{\epsilon}\eta^{3}}{3(e^{\epsilon}-1)(\eta-1)^{2}}.
\end{equation}

On the other hand, $I_{1}(\eta)$ is not always larger than zero for an arbitrary parameter $\eta>1$, which means that we have no ability to establish optimal I-PTT than Duchi's scheme. In a word, we finish the proof of Theorem 8. \qed

\end{document}